\documentclass[preprint,aps]{revtex4}
\usepackage{epsfig}
\usepackage{latexsym}
\usepackage{amssymb}
\usepackage{wasysym}
\usepackage{pifont}
\usepackage{graphicx}
\usepackage[dvips]{color}

\begin{document}

\title{Multifractality in Rotational Percolation Models}
 
\author {Santanu Sinha and S. B. Santra}

\affiliation{Department of Physics, Indian Institute of Technology
  Guwahati, Guwahati-781039, Assam, India.}

\date \today

\begin{abstract}
In rotationally constrained percolation models, a site of a
percolation cluster could be occupied more than once from different
directions due to the nature of the rotational constraint. A state
variable $s_i$ is assigned to each lattice site whose value corresponds
to the number times it has been visited during the growth of a
cluster. It is proposed here that the percolation transition and the
multifractal aspects of infinite percolation clusters under rotational
constraint can be studied defining suitable measures in terms of the
state variable $s_i$.  This method does not require to introduce any
external agency like an electric current or a random walker in order
to explore multifractality as in the case of ordinary percolation. The
state variable representation also describes the universality class of
the percolation models appropriately. 
\end{abstract}

\maketitle

\section{Introduction}
Multifractals appear in a wide range of situations like energy
dissipation in turbulent flows\cite{turbl}, electronic eigenstates at
metal insulator transition\cite{mit}, diffusion in porous
structures\cite{sapo}, diffusion limited aggregation\cite{jensen},
fluctuations in finance\cite{finc}, dynamics of human
heartbeat\cite{hhb} and many others. The multifractal properties
associated with the infinite percolation clusters at the percolation
threshold $p_c$ is considered in this paper. In ordinary percolation
(OP)\cite{bunde}, a cluster is generated by occupying a lattice site
randomly with a probability $p$ or remains empty with a probability
$(1-p)$. Each site of a percolation cluster then has two states,
occupied or empty. The average $q$ moments of the cluster size
distribution of percolation clusters are linearly dependent on the
moment $q$ and described by a single fractal dimension
$d_f$\cite{prc}. In order to study the multifractal aspects of
percolation clusters, usually, a current
distribution\cite{aharony,janssen} or a random walker\cite{havlin} is
introduced. However, in the presence of rotational constraint on the
percolation model, an occupied site has always a direction associated
with it and a site can be re-occupied from different directions. A
site can be occupied at most $z$ times from $z$ different directions
on a given lattice with coordination number $z$. There are two such
well studied rotational percolation models exists, spiral percolation
SP\cite{rsb} and directed spiral percolation\cite{dsp}. In these
models, a state variable $s_i$ then can be assigned to each lattice
site and whose value will correspond to the number of times a site is
visited during the growth of the cluster. The value of $s_i$ then can
change from $0$ to $4$ on a square lattice and $0$ to $6$ on a
triangular lattice for SP and DSP models whereas it has only two
states $1$ and $0$ in case of OP. In this paper, a new methodology is
proposed to study the percolation transition in terms of the state
variable $s_i$. Studying the physical properties associated with the
state variable $s_i$, the percolation transition is possible to
establish at the same percolation threshold $p_c$ defined
geometrically. The spanning clusters at $p=p_c$ are
fractal. Distribution of $s_i$ on the fractal substrate is very
similar to mass distribution on a geometrical support usually taken
for multifractal study\cite{feder}. In order to explore the
multifractal aspects of the spanning percolation clusters in these
rotationally constrained percolation models at $p=p_c$, it is now
possible to define a suitable multifractal measure in terms of
$s_i$. In this way, one does not need to introduce any other external
agency like electric current or random walker in the model as it is
usually done in the case of OP clusters. The variable $s_i$ is inbuilt
in the rotational models and represents an inherent property of SP and
DSP. It is found that the exponents associated with the $q$ moments of
the measure defined in terms of $s_i$ are not limited by any linear
dependence on the moment $q$ for the rotational percolation models. It
then indicates that the measure has multifractal character.

Below, description of the rotational percolation models will be given
and the multifractal aspects of the spanning clusters will be
investigated.

\section{Rotational Percolation Models}
There are two well studied rotationally constrained percolation
models, spiral percolation (SP)\cite{rsb} and directed spiral
percolation (DSP)\cite{dsp}. In these models, clusters are grown
following single cluster growth Monte Carlo (MC) algorithms. In these
algorithms, the central site of the lattice is occupied with unit
probability. The nearest neighbors of the central site is occupied
with equal probability $p$ in the first time step. As soon as a site
is occupied, the direction from which it was occupied is assigned to
it. Lists of eligible sites for occupation in the next MC time steps
are identified and they are occupied with probability $p$. In these
algorithms, an occupied site can be reoccupied from a different
direction but it is forbidden for occupation from the same
direction. Once a site is rejected for occupation it is forbidden for
occupation throughout the simulation from any possible direction. Each
MC time step can be considered as parallel update of nearest
neighbours of already occupied sites. The growth of a cluster stops if
there is no eligible site available for occupation. Identification of
eligible sites for occupation in a MC step for SP and DSP are given
below.

\begin{figure}
\centerline{\hfill 
\psfig{file=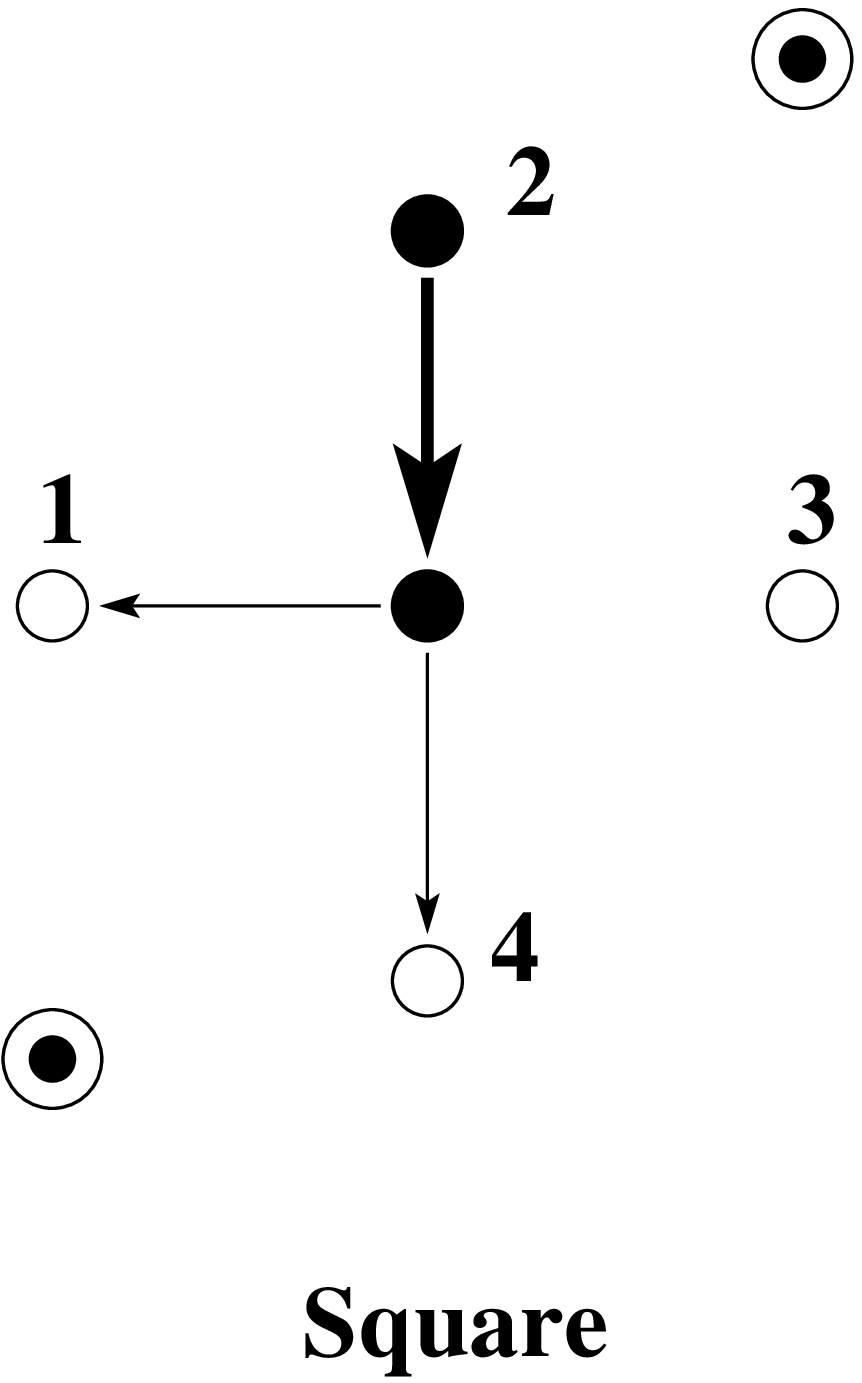,scale=0.45}
\hfill
\psfig{file=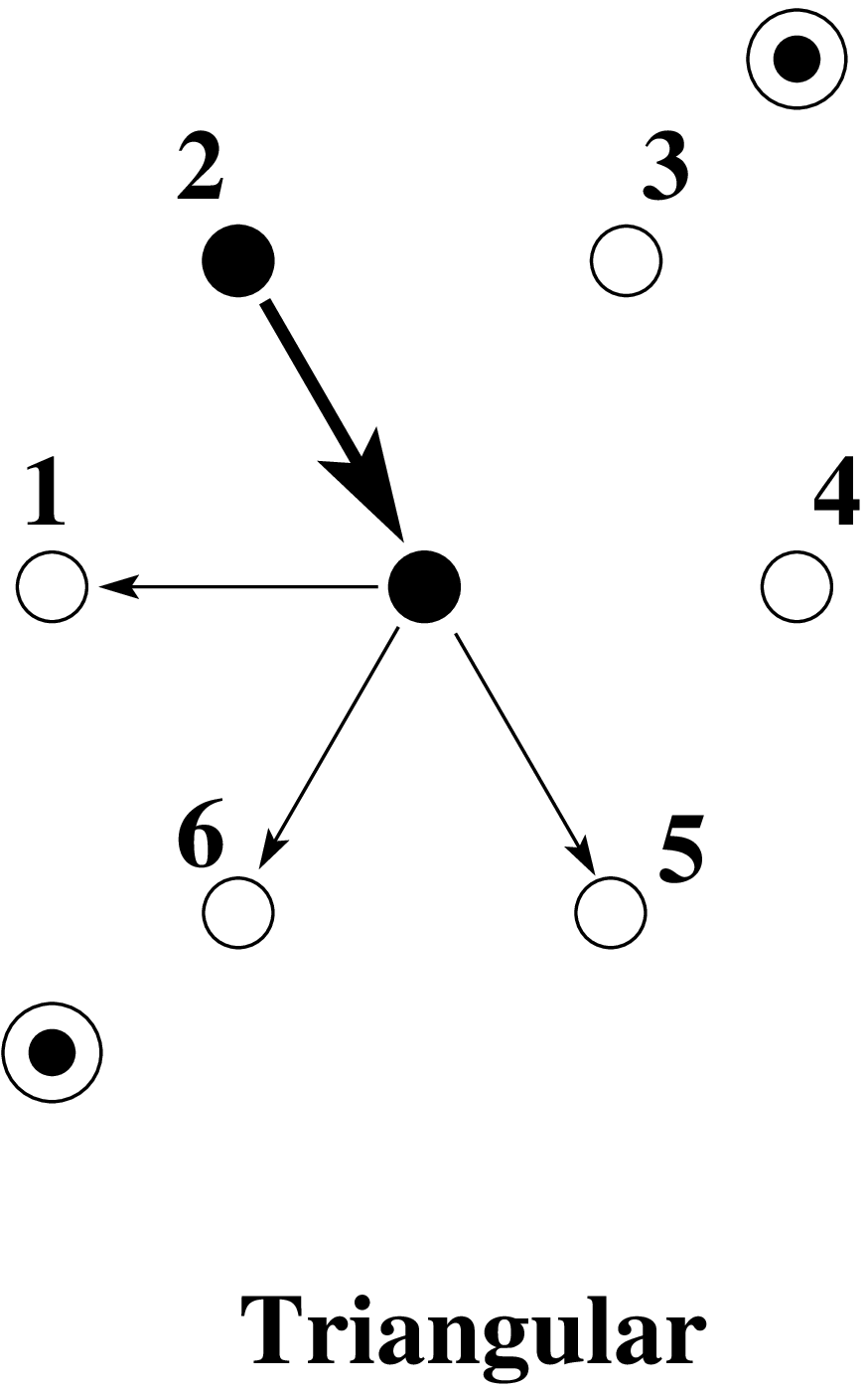,scale=0.45}\hfill}
\medskip
\centerline{\hfill ($a$) SP model \hfill}
\bigskip
\bigskip
\centerline{\hfill
\psfig{file=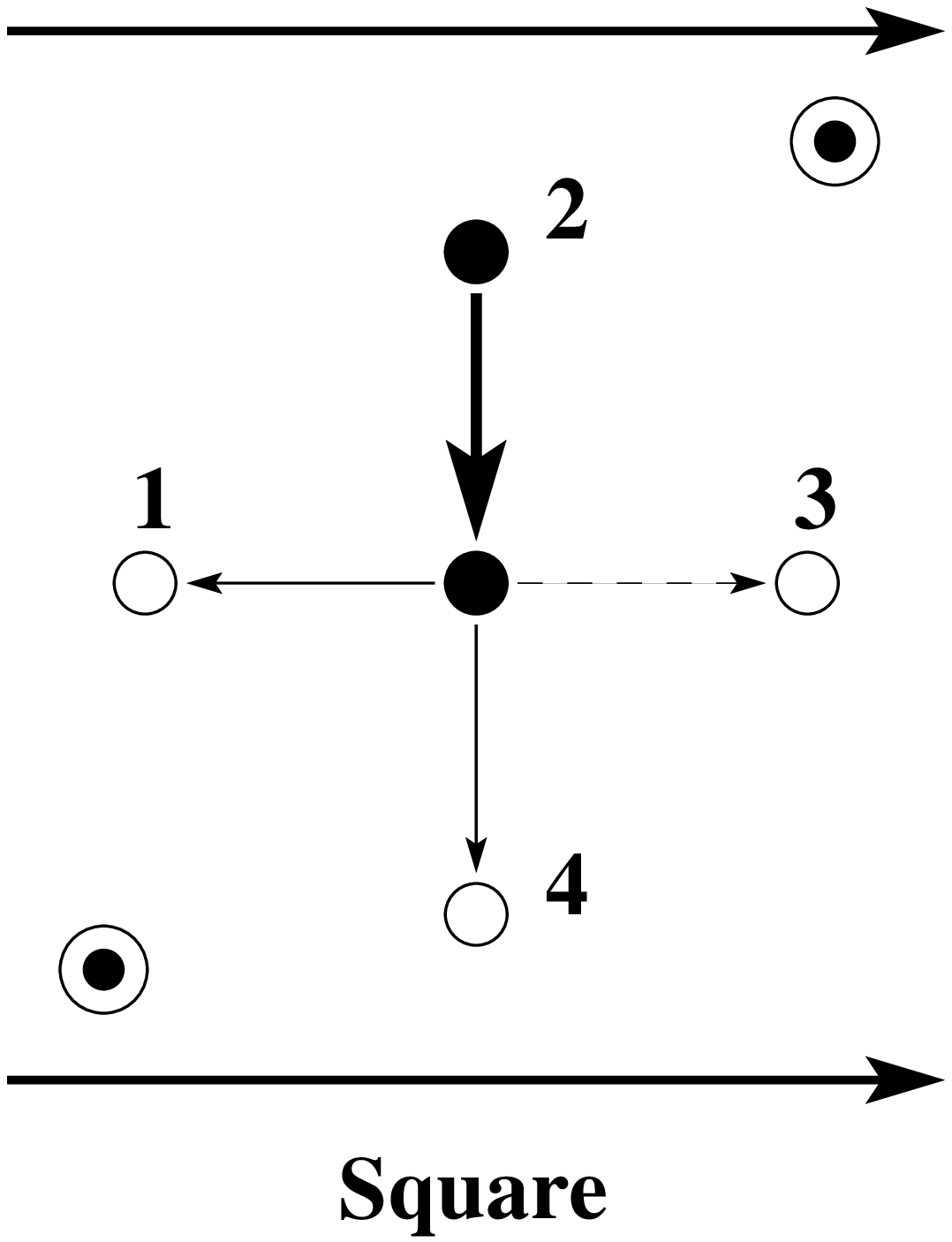,scale=0.45}
\hfill
\psfig{file=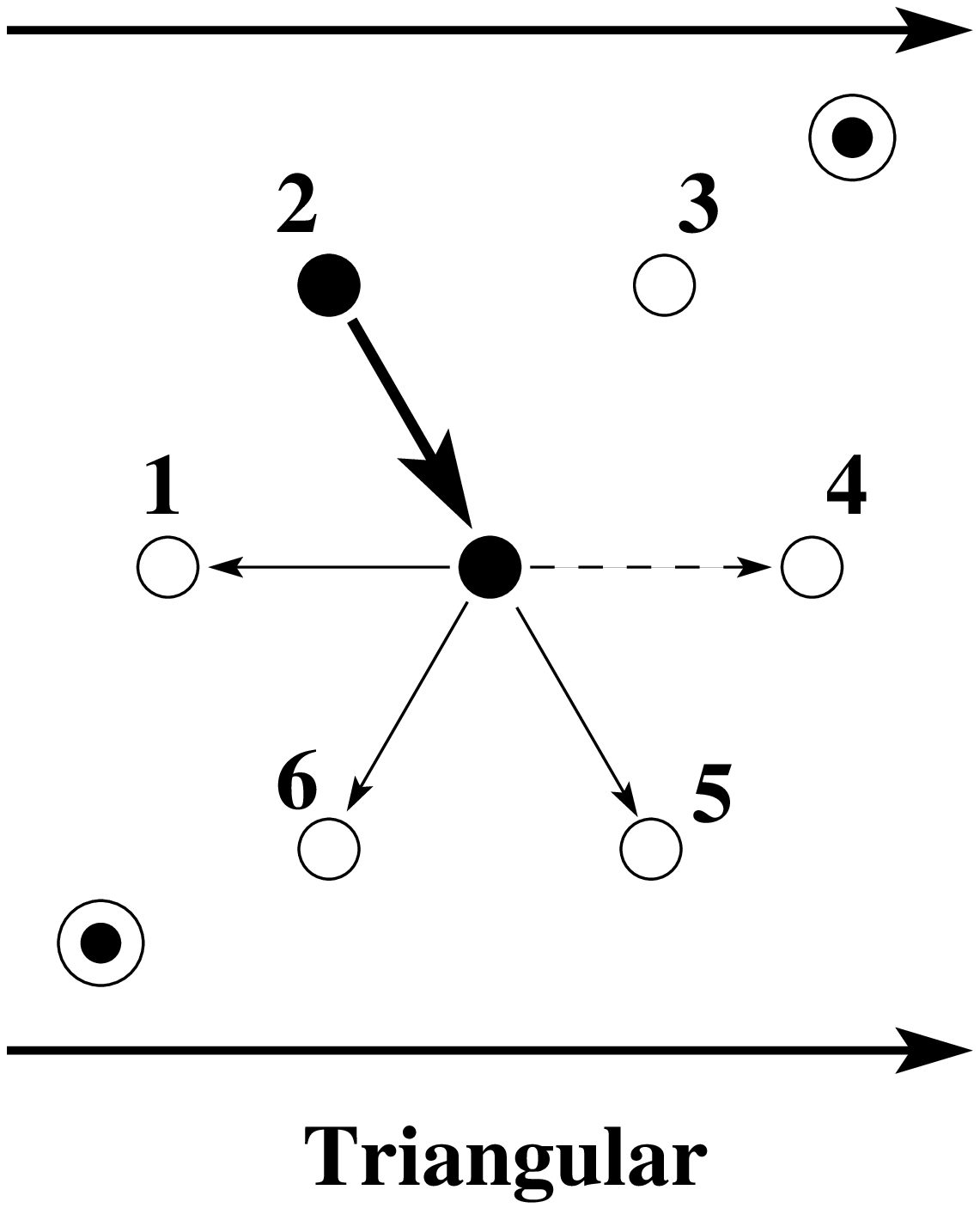,scale=0.45}\hfill}
\medskip
\centerline{\hfill  ($b$) DSP model \hfill}
\smallskip
\caption{\label{demo} Selection of empty nearest neighbours in a MC
  step in ($a$) SP model and ($b$) DSP model on the square and
  triangular lattices. Black circles are the occupied sites and the
  open circles are the empty sites. Thick long arrows from left to
  right represent directional constraint ($E$). The clockwise
  rotational constraint ($B$) is shown by encircled dots. The central
  site here is occupied from site $2$ and shown by short thick
  arrows. The eligible sites for occupation due to $E$ field are shown
  by dotted arrows and thin solid arrows indicate the same due to $B$
  field on both the lattices. }
\end{figure}

In SP, only a rotational field $B$ is applied perpendicular to the
plane ($xy$) of the lattice and directed along the negative $z$-axis
(of a right handed coordinate system). As an effect of the $B$ field,
empty sites in the forward direction and in the clockwisely rotational
direction are eligible for occupation. The forward direction is the
direction from which the present site is occupied and the sense of
rotational direction is defined with respect to the forward
direction. The selection of eligible sites for occupation in SP model
is demonstrated in Fig.\ref{demo}($a$) for both the square and
triangular lattices. There are two eligible sites on the square
lattice and three eligible sites on the triangular lattice due to
higher coordination number. The eligible sites are then occupied with
probability $p$ and the clusters grow isotropically on the
lattice\cite{rsb}.

In case of DSP, a crossed directional field $E$ is also applied in
addition to the rotational field $B$. The $E$ field is applied from
left to right in the plane of the lattice and $B$ is, as usual,
directed along the negative $z$ axis. Due to $E$ field, empty site on
the right of an occupied site is eligible for occupation along with
the eligible sites for occupation due to $B$ field. The selection of
eligible sites for occupation in DSP model is demonstrated in
Fig.\ref{demo}($b$) for both the square and triangular lattices. The
components of the directional constraint remain the same on both the
lattices whereas there is an extra component of rotational constraint
on the triangular lattice as in SP. The eligible sites are then
occupied with probability $p$. Once a site is rejected for occupation
it is forbidden for occupation throughout the simulation from any
possible direction due to both $E$ and $B$ fields. Because of the
simultaneous presence of both the $E$ and $B$ fields crossed to each
other, a Hall field $E_H$ appears in the system perpendicular to both
$E$ and $B$. As a result, an effective directional constraint $E_{\sf
eff}$ acts on the system along the diagonal from left upper to right
lower corner of the lattice. Here, the clusters grow anisotropically
along the effective field $E_{\sf eff}$\cite{dsp}.

The coordinate of an occupied site in a cluster is denoted by
$(x$,$y)$. Periodic boundary conditions are applied in both directions
and the coordinates of the occupied sites are adjusted accordingly
whenever the boundary is crossed. At each time step the span of the
cluster in the $x$ and $y$ directions $L_x = x_{max} - x_{min}$ and
$L_y = y_{max} - y_{min}$ are determined. If $L_x$ or $L_y\ge L$, the
system size, then the cluster is considered to be a spanning cluster.

Since an occupied site can be reoccupied from different directions due
to the presence of rotational constraint in both the models, a site
then can be occupied at most $z$ times from $z$ possible directions on
a given lattice where $z$ is the coordination number of the
lattice. The value of $z$ is $4$ on the square lattice and it is $6$
on the triangular lattice. This is unlike in the case of the ordinary
percolation where a site is occupied only once. It is an essential and
a special feature of SP and DSP models. It is thus possible to assign
a state variable $s_i$ to each site and the value of $s_i$ will
represent the number of times a site is occupied or visited. Initially
the values of $s_i$ are all set to zero. As soon as a site is occupied
from any direction, the value of $s_i$ is increased by $1$. The values
of the state variable is then given by $s_i=0,\cdots,z$ on a given
lattice. $s_i=0$ corresponds to unoccupied site. The state variable
$s_i$ here is similar to the Ising spins with $(z+1)$ states. In case
of OP, $s_i$ could have values only $0$ and $1$ corresponding to
unoccupied and occupied sites.

Percolation clusters are generated here in the presence of rotational
constraint on the square and triangular lattices of size $L = 1024$
for both SP and DSP models. The percolation transition will be
established first at the original $p_c$ by calculating ``spontaneous
magnetization'' in terms of $s_i$. The multifractal aspects of the
measure distribution on the spanning clusters, defined in terms of
$s_i$, are then investigated for both isotropic SP and anisotropic DSP
clusters at $p=p_c$. The cluster properties are averaged over $5\times
10^4$ spanning clusters.

\section{Percolation threshold and Spanning Clusters}

\begin{figure}
\centerline{\hfill
  \psfig{file=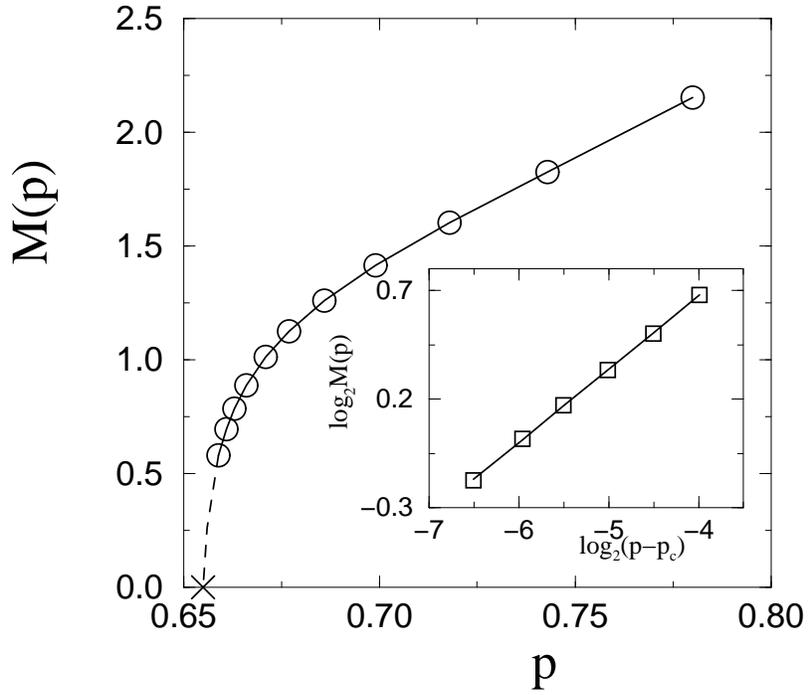,width=0.65\textwidth}\hfill} 
\smallskip
\caption{\label{orderp} Plot of spontaneous magnetization $M(p)$
  against $p$ for DSP model defined on a square lattice of size
  $L=1024$. Percolation threshold $p_c$ is marked by a cross on the
  $p$ axis. In the inset, $M$ is plotted with ($p-p_c$). From the
  slope, the exponent $\beta$ is obtained as $\approx 0.32$.}
\end{figure}

Geometrically, the critical percolation probability $p_c$ is the
maximum probability below which no spanning cluster appears.  At
$p=p_c$, a spanning cluster appears for the first time in the
system. In single cluster growth approach, the threshold $p_c$ is
generally identified by measuring the probability to have a spanning
cluster ($P_\infty$) as a function of $p$, the occupation
probability. $P_\infty$ goes to zero continuously as $p$ tends to
$p_c$ from above. In the state variable formalism, the value of $p_c$
can be identified in terms of ``spontaneous magnetization'' $M(p)$
defined in terms of the state variable $s_i$. $M(p)$ is defined as
\begin{equation}
\label{opm}
M(p) = \frac{1}{N_{tot}}\sum_{j=1}^{N_{tot}} m_j(p),
\hspace{1cm} m_j(p) = \frac{1}{L^2}\sum_{i=1}^{L^2} s_i(p)
\end{equation}
where $L$ is the lattice size and $N_{tot}$ is the total number of
spanning clusters generated. $m_j(p)$ represents the magnetization per
site for the $j$th spanning cluster generated using single cluster
growth method. At $p=1$, all the sites of an infinite cluster are
expected to be occupied $z$ times where $z$ is the coordination number
of the lattice and the size of the infinite cluster will be of the
order of $L^2$, square of the system size. Thus, $M(1)$ is equal to
$z$. As $p$ tends to the percolation threshold $p_c$ from above,
$M(p)$ is expected to go to zero continuously from its maximum value
$z$ at $p=1$ not only because the sites will be occupied less number
of times but also the spanning cluster will disappear. $M(p)$ is
measured on the square lattice ($z=4$) for DSP model and it is plotted
against $p$ in Fig.\ref{orderp}. It can be seen that it is going to
zero at $p=p_c$ as expected and the value of $p_c$ is the same as that
determined by geometrical approach $p_c=0.6550\pm
0.0005$\cite{dsp}. It is also expected that $M(p)$ becomes singular at
$p=p_c$ with an exponent $\beta$ as $M(p) \approx (p-p_c)^\beta$. In
the inset of Fig.\ref{orderp}, the power law has been verified and the
exponent $\beta$ is determined approximately as $\beta\approx 0.32$,
close to the already obtained value ($0.31\pm0.01$)\cite{dsp}. The
state variable formalism of the rotationally constrained percolation
models is then consistent with that of the usual geometrical
approach. The value of $p_c$ has also been recovered within error bar
in the case of SP. Other critical properties of the models can also be
identified in terms of the state variable $s_i$ and a scaling theory
is possible to develop.

\begin{figure}
\centerline{\hfill \psfig{file=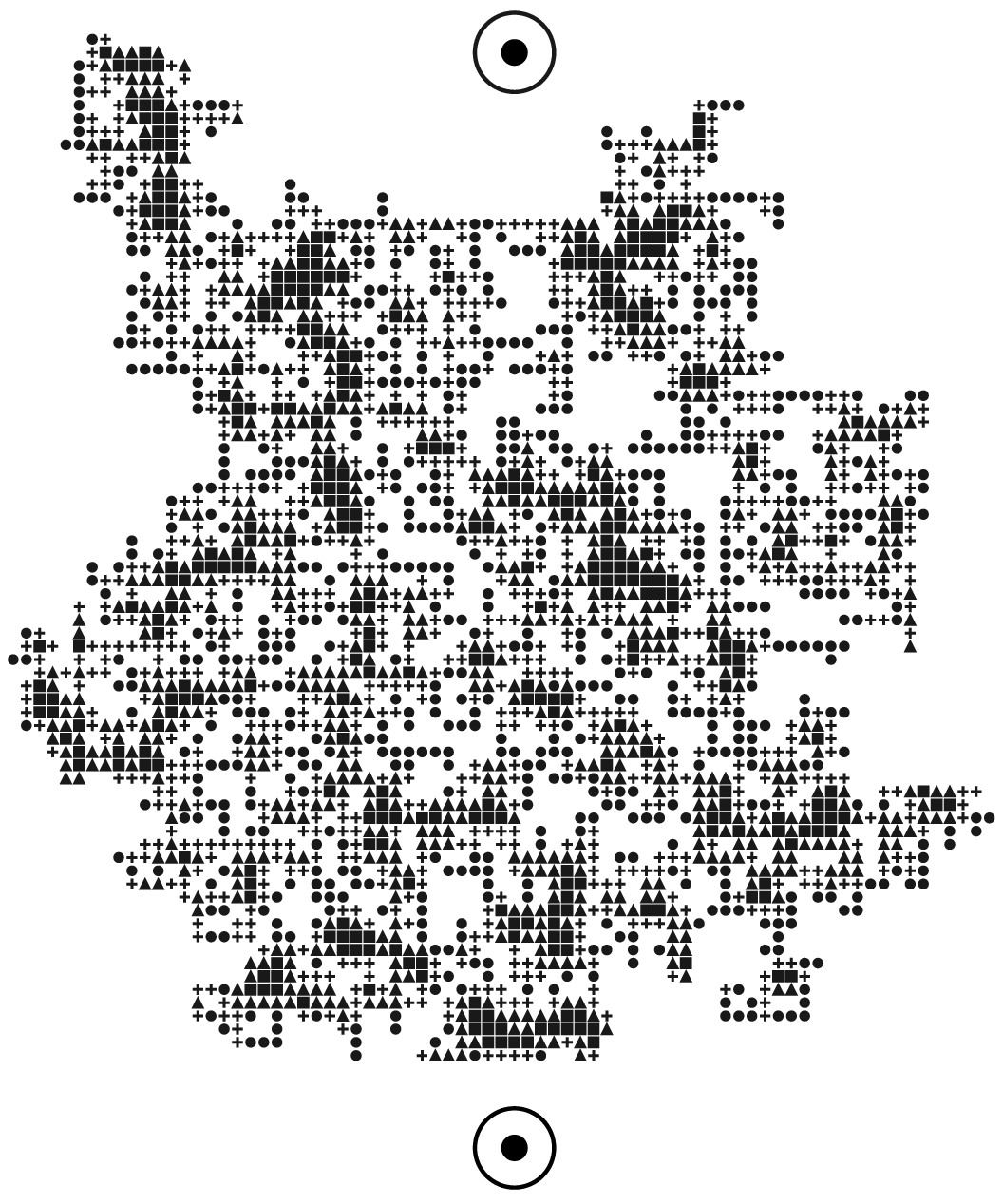,scale=0.70}
\hfill\hfill \psfig{file=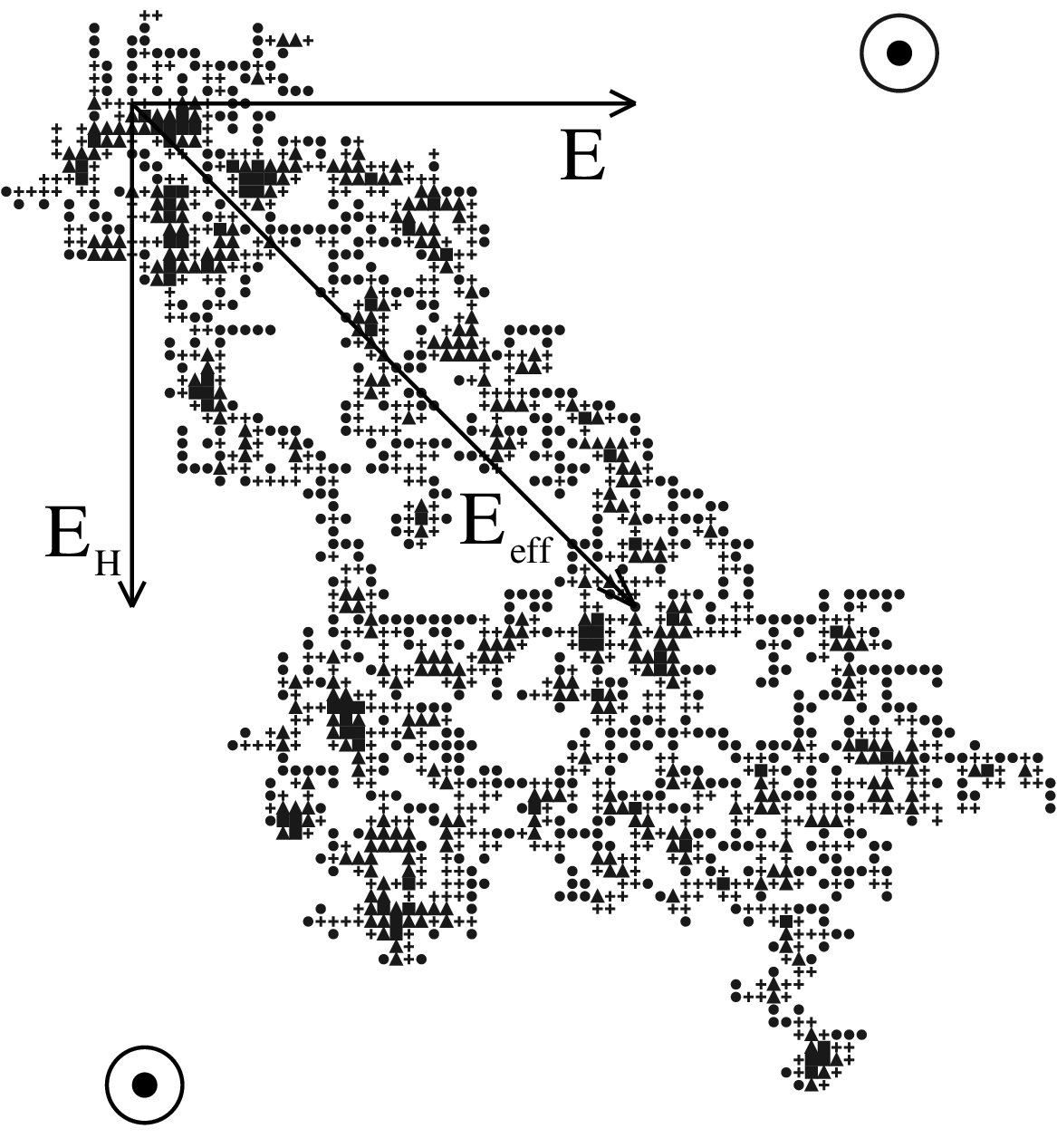,scale=0.70}\hfill}
\medskip
\centerline{\hfill ($a$) \hfill\hfill  ($b$) \hfill}
\smallskip
\caption{\label{cluster} Typical spanning clusters at $p=p_c$ on the
  square lattice of size $L=2^6$ ($a$) for SP and ($b$) for DSP
  models. The encircled dots represent the rotational field $B$ and
  the arrows represent the directional field $E$. Different symbols in
  the clusters represent different values of $s_i$ as ($\bullet$) for
  $s_i=1$, ($+$) for $s_i=2$, ($\blacktriangle$) for $s_i=3$ and
  ($\blacksquare$) for $s_i=4$. The empty white space represents
  $s_i=0$. It can be seen that the state variable is randomly
  distributed over the fractal spanning clusters.}
\end{figure}

Typical spanning clusters at $p=p_c$ generated on the square lattice
of size $L=2^6$ are shown in Fig.\ref{cluster}($a$) for SP and in
Fig.\ref{cluster}($b$) for DSP. Different values of the state variable
$s_i$ is represented by different symbols as $s_i=0$ (white space),
$s_i=1$ (filled circle), $s_i=2$ (plus), $s_i=3$ (filled triangle) and
$s_i=4$ (filled square). It can be seen that not only the state
variables $s_i$ are randomly distribution of over the spanning cluster
but also the higher states form small islands allover the spanning
cluster. This is similar to the situation of mass distribution on a
geometrical support generally taken for multifractal
study\cite{feder}. However, the state distribution over the spanning
cluster is not a simple iterative process of mass distribution over a
geometrical support. The spanning clusters consist of subsets of sites
occupied once, twice upto a subset of sites occupied $z$ times where
$z$ is the coordination number of the lattice considered. A particular
subset may appear several times in a spanning cluster at different
stages of growth of the cluster during a large number of MC steps. A
multiplicative cascade of these subsets is then formed in a
complicated manner during the growth of the cluster. The $s_i$
distribution on the spanning cluster is then expected to have many
folds. It is then interesting to investigate the moments of the $s_i$
distribution over fractal objects, the spanning clusters here. It
could also be noted that the SP cluster is compact and isotropic but
the DSP cluster is highly rarefied and anisotropic. The elongation of
the DSP cluster is along the effective field $E_{\sf eff}$ appeared in
the system.  However, the clusters are not merely DP clusters along
$E_{\sf eff}$. It has already been found that both SP and DSP belong
to new universality classes than that of DP or OP\cite{rsb,dsp}. The
fractal dimension $d_f$ of the spanning clusters were found as $d_f
\approx 1.733$\cite{dsp} for DSP and $\approx 1.957$\cite{rsb} for
SP. Geometrical properties of the percolation clusters are governed by
this single exponent $d_f$. However, in the following it will be
demonstrated that a measure defined in terms of the state variable
$s_i$ is not restricted by a single exponent rather needs a sequence
of fractal dimensions to characterize the measure.

\section{Mulifractality}
\begin{figure}
\centerline{\hfill  
  \psfig{file=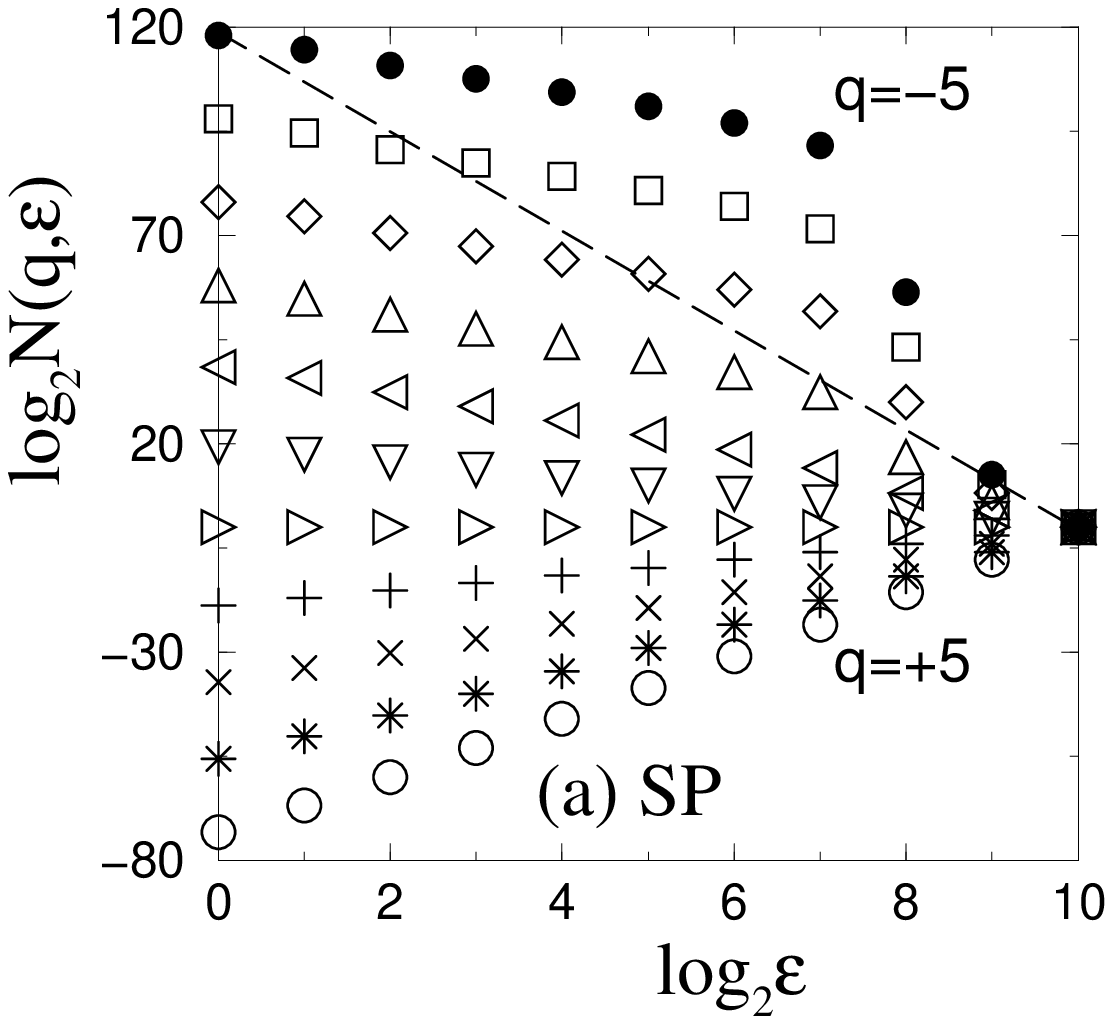,width=0.5\textwidth}\hfill
  \psfig{file=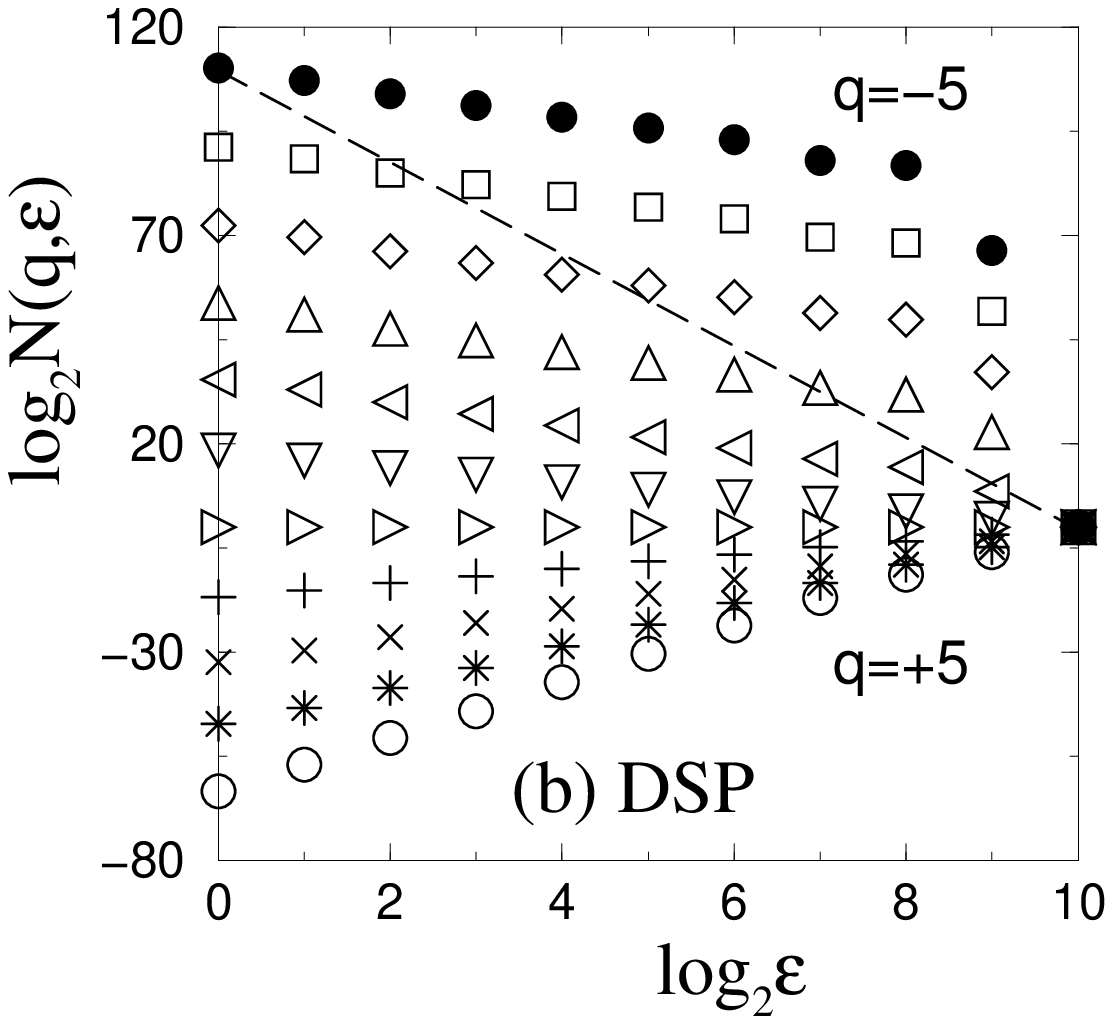,width=0.5\textwidth}\hfill}
\smallskip
\caption{\label{tauq1} Plot of weighted number of box $N(q,\epsilon)$
  versus the box size $\epsilon$ for $q=-5$ to $q=5$ in step of $1$
  for SP in ($a$) and for DSP in ($b$) for the spanning clusters
  generated on the square lattice of size $L=1024$. The symbols are:
  $(\bullet)$ for $q=-5$, $(\Box)$ for $q=-4$, $(\diamond)$ for
  $q=-3$, $(\triangle)$ for $q=-2$, $(\triangleleft)$ for $q=-1$,
  $(\triangledown)$ for $q=0$, $(\triangleright)$ for $q=1$, $(+)$ for
  $q=2$, $(\times)$ for $q=3$, $(*)$ for $q=4$, and $(\circ)$ for
  $q=5$. It can be seen that box counting method is not working for
  $q<0$. The expected behaviour is shown by dashed lines for $q=-5$ in
  both ($a$) and ($b$).}
\end{figure}

In order to study multifractality a suitable measure has to be
defined. In general, a multifractal measure is related to the
distribution of a physical quantity on a geometrical support
\cite{feder}. The geometrical support here is the spanning percolation
cluster at $p=p_c$ for the rotationally constrained percolation
models. The distribution of the relative probability of a state over
fractal substrates is a possible multifractal measure here. It is
similar to the mass distribution on a geometrical support. The
multifractal measure $\mu_i$ is then defined as
\begin{equation}
\label{mesr}
\mu_i=s_i/\sum_{i=1}^{L^2}s_i
\end{equation}
where $s_i$ is the state variable associated with each lattice
site. $\mu_i$ can be called as relative state variable. Note that the
measure $\mu_i$ is normalized to unity when summed over the whole
lattice. The maximum value of the measure is $\mu_{max} = z/\sum s_i$
and the minimum non-zero measure is $\mu_{min} = 1/\sum s_i$ where $z$
is the coordination number of the lattice. To obtain the multifractal
nature of the distribution $\mu_i$, it is necessary to study the
scaling of the $q$-moments of the measure over different length scales
on the geometrical support. If the measure $\mu_i$ is multifractal and
the support is divided into $n_\epsilon$ boxes of size $\epsilon$,
then the weighted number of boxes $N(q$,$\epsilon)$ is given by
\begin{equation}
\label{neq}
N(q,\epsilon)=\sum_{j=1}^{n_\epsilon} \mu_j^q \approx \epsilon^{-\tau(q)}
\end{equation}
where $\mu_j$ is the sum of the relative state variable in the $j$th
box. Here $\tau(q)$ can be called as ``state exponent''. The weighted
number of box $N(q,\epsilon)$ is determined as a function of the box
size $\epsilon$ using box counting method for a given $q$. The boxes
with at least one occupied site are only considered. The weighted
number of boxes $N(q,\epsilon)$ are plotted against the box size
$\epsilon$ for $q=-5$ to $q=+5$ for SP in Fig.\ref{tauq1}($a$) and for
DSP in Fig.\ref{tauq1}($b$) generating spanning clusters on the square
lattice of size $L=1024$. It can be seen that the slopes of the plots
change continuously for positive $q$ up to $q=0$. For $q<0$, it seems
that the usual box counting method adopted here is not working. The
values of $N(q,\epsilon)$ remain unchanged over several box sizes
$\epsilon$ starting from the smallest box size for a given $q$ in both
the models. It is expected that the plot should follow a straight line
passing through the points at $\epsilon = 1$ and $\epsilon = 2^{10}$,
the system size, in log-log scale since these two extreme points are
not effected by the box size. It is shown by dashed lines for $q=-5$
in both the plots. It is observed that the value of $N(q,\epsilon)$
jumps suddenly when the box size is reduced less than one quarter of
the system size. This is due to the fact that at this box size there
is at least one box appearing with a small measure and consequently
the sum in Eq.\ref{neq} diverges due to negative moment. The
appearance of large box sizes with small measures is because of the
fact that the spanning percolation clusters contain holes of all
possible sizes. Difficulties in determining weighted number of boxes
for $q<0$ for the measure distribution on random structures are
already reported in the literature\cite{mfbxnq}. The slopes of the
plots in Fig.\ref{tauq1} also remain almost unchanged with the moment
$q$ for $q<0$. The weighted number of boxes has increased
proportionally with higher negative moments. It has been verified that
the estimation of $\tau(q)$ by fitting only through the smaller box
sizes leads to a discontinuity in the plot of $\tau(q)$ versus $q$
which is expected to be a smooth function of $q$. Discontinuity in the
plot of $\tau(q)$ versus $q$ was also observed in the cases of
resistance fluctuations in randomly diluted networks \cite{aharony}
and in diffusion limited aggregation \cite{stanley}. In these cases,
there are breakdown of multifractal characters for negative moments
due to exponential decay of the smallest measures.

\begin{figure}
\centerline{\hfill
  \psfig{file=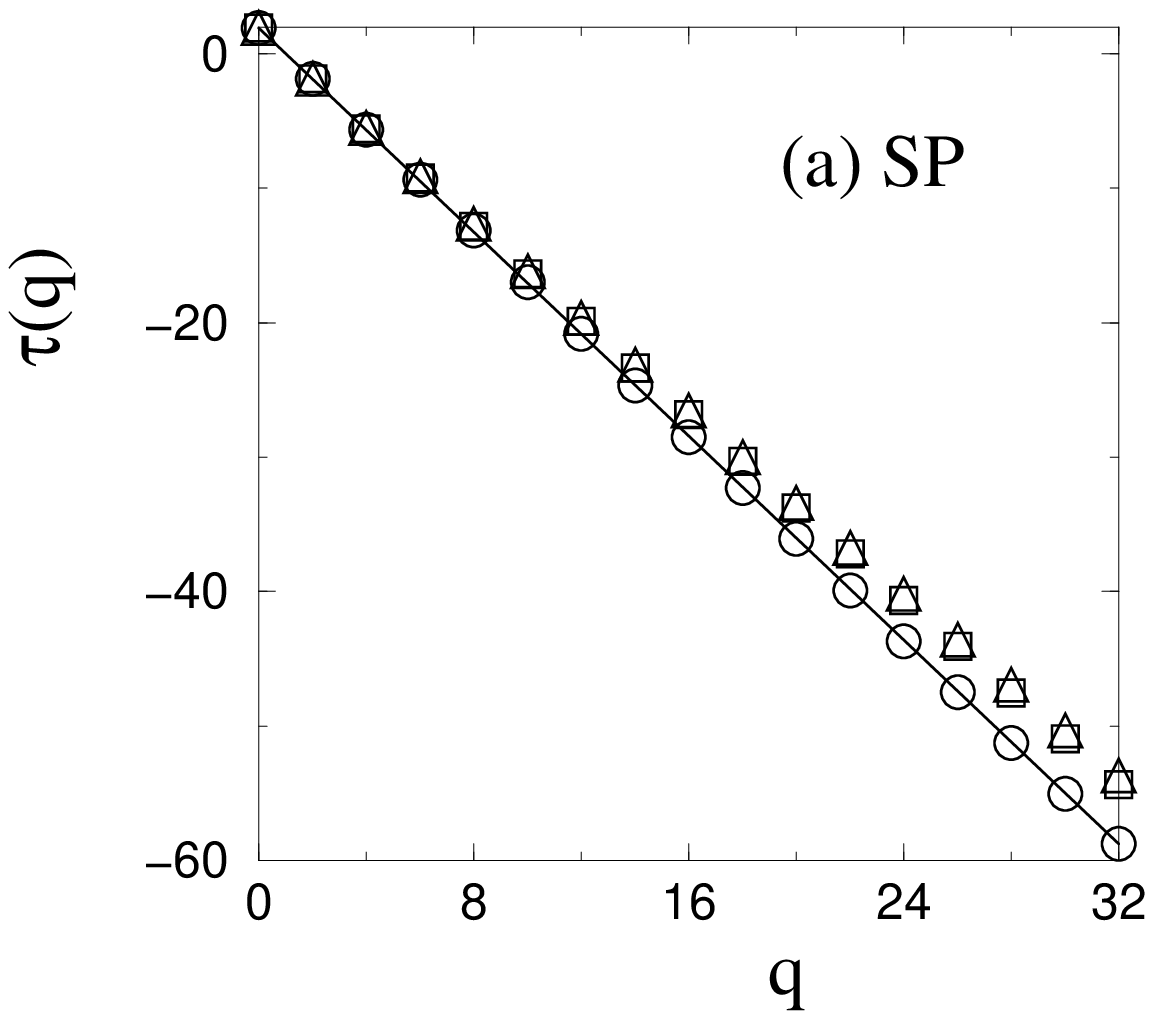,width=0.5\textwidth}\hfill\hfill
  \psfig{file=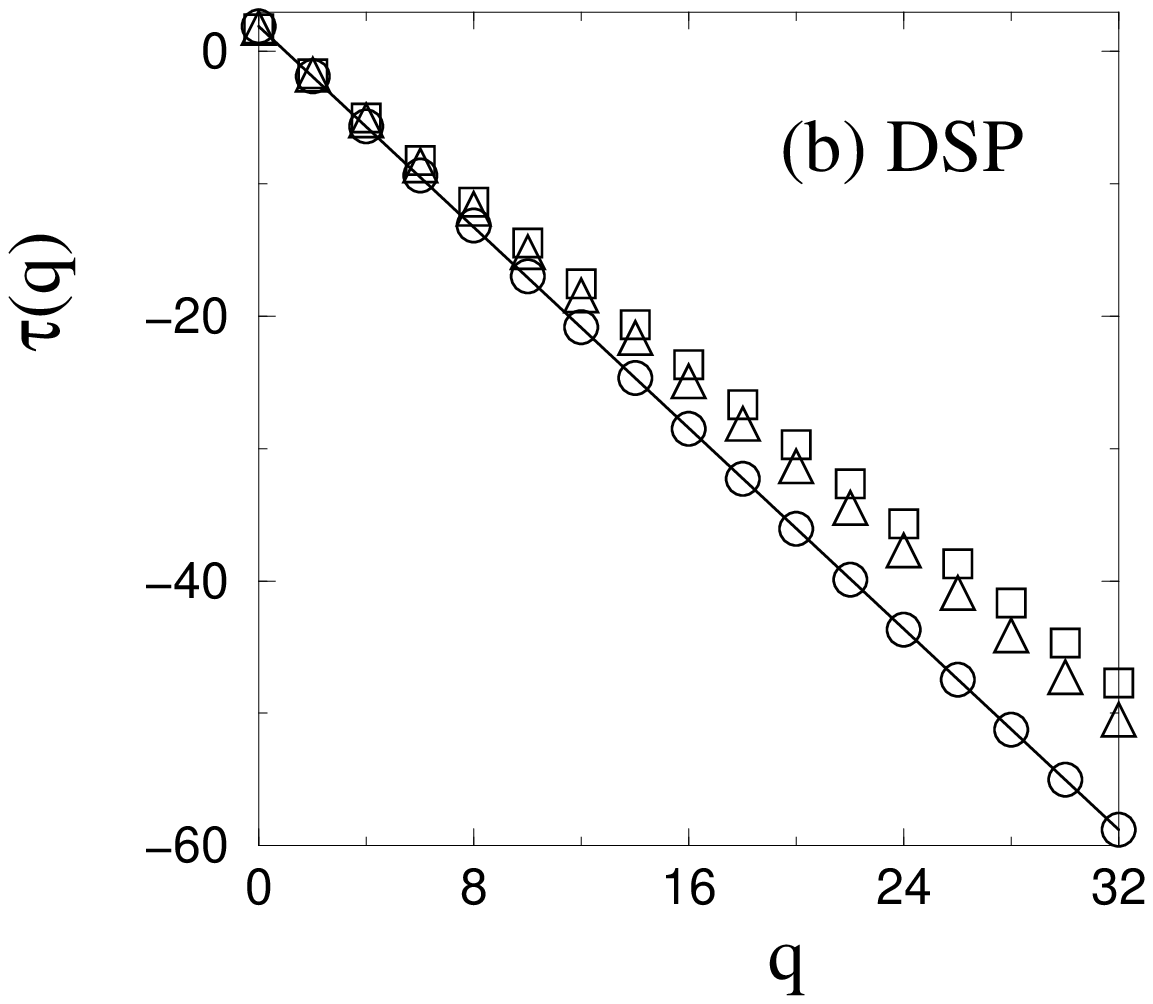,width=0.5\textwidth}\hfill }
\smallskip
\caption{\label{tauq2} Plot of the state exponent $\tau(q)$ versus the
  moment $q$ for SP in ($a$) and for DSP in ($b$) for $q\ge
  0$. Squares represent the square lattice data and triangles
  represent triangular lattice data respectively. Circles represent
  data of OP spanning clusters on a square lattice. The solid straight
  line represents the linear dependence of $\tau(q)$ on $q$ expressed
  in Eq.\ref{tqdf} for OP. The measured values of $\tau(q)$ for OP
  follows the straight line behaviour. For DSP and SP, $\tau(q)$ has a
  non-linear dependence on the moment $q$.}
\end{figure}

Multifractal characteristics of the spanning clusters of rotationally
constrained percolation models are then analyzed here taking large
positive moments, changing $q$ from $0$ to $32$. Analysis has been
made on the square and triangular lattices of size $L=1024$ for both
SP and DSP models and the results are compared with that of the OP
model. In Fig.\ref{tauq2}, $\tau(q)$ is plotted against $q$, ($a$) for
SP model and ($b$)for DSP model. In both the plots the squares
represent the square lattice data and the triangles represent the
triangular lattice data. Circles represent the data obtained for OP
model. It is found that $\tau(0)$ is $\approx d_f$, the fractal
dimension of the corresponding spanning clusters and $\tau(1)$ is
$\approx 0$ here for all three models, OP, SP and DSP. $\tau(0)$
corresponds to the dimension of the support which are the spanning
percolation clusters of different models considered here and $\tau(1)$
is zero because $\sum_i\mu_i = 1$. It is interesting to notice that
the values of $\tau(q)$ for DSP and SP model depend on the moment $q$
in a nonlinear way for positive moments. If the measure $\mu_i$ is
characterized by a single fractal dimension $d_f$, the state exponent
$\tau(q)$ should have a constant gap between two consecutive
exponents\cite{prc} and consequently should have a linear dependence
on $q$. In that case, a relationship between $\tau(q)$ and $q$ in
terms of $d_f$ can be obtained as
\begin{equation}
\label{tqdf}
\tau(q) = -(q-1)d_f.
\end{equation}
This relation is shown in Fig.\ref{tauq2} by a solid line for OP
taking $d_f=91/48\approx 1.896$. The values of $\tau(q)$ obtained
numerically for OP (circles) considering a two state model for which
$s_i=0$ and $1$ are in good agreement with with Eq.\ref{tqdf}. There
are few more things to notice. First, the state exponents of OP obey
the constant gap equation given in Eq.\ref{tqdf} as expected. The
constant gap scaling was also observed for the mean number of distinct
sites visited by a random walker on spanning percolation cluster by
Murthy {\em et al}\cite{murthy}. As a consequence, the measure
distribution is mono-fractal. Second, the values of $\tau(q)$ for SP
and DSP are deviated form straight line behaviour and have non-linear
dependence on the moment $q$. Thus, each moment of the measure $\mu_i$
needs a new exponent to characterize in these models. Third, the
functional dependence of $\tau(q)$ on $q$ is found different for all
three models. This is expected because the models, OP, SP and DSP,
belong to different universality classes. Fourth, the $\tau(q)$ values
on the square and triangular lattices are almost the same for the SP
model whereas they are considerably different for the DSP model. This
is also in agreement with the fact that the critical properties hold
universality in the SP model whereas there is a breakdown of
universality in DSP model between the square and triangular lattices
in two dimensions\cite{sinha}. Finally, the fact that a sequence of
exponents is required to characterize the moments of the measure
confirms the multifractal nature of $\mu_i$ distribution in SP and DSP
models. It should be noted here that in SP and DSP the spanning
clusters consist of four or six subsets depending on the number of
nearest neighbours on a given lattice. The values of $\tau(q)$ then
may be possible to obtain in terms of the fractal dimensions of the
subsets consisting the spanning cluster coupled with a nonlinear
dependence on $q$. However, it is difficult to determine the fractal
dimensions of the individual subsets as they are generally small
isolated islands in a spanning cluster as well as the nonlinear nature
of $\tau(q)$.

\begin{figure}
  \centerline{\hfill \psfig{file=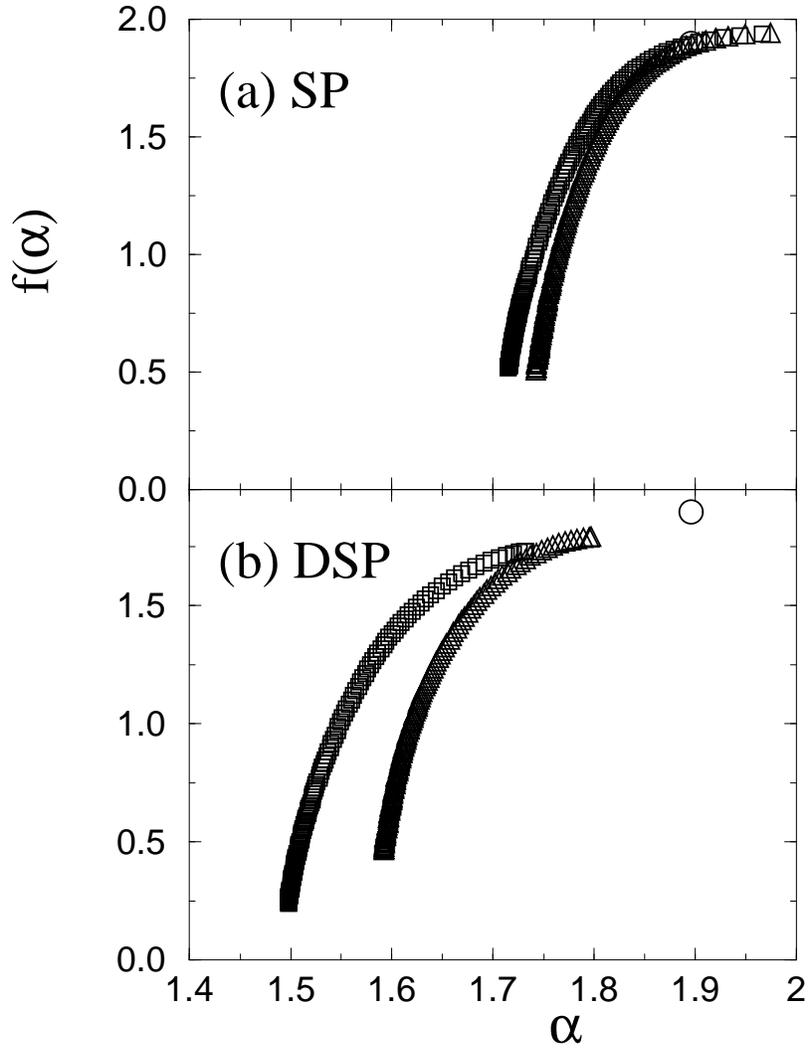,width=0.65\textwidth}\hfill}
\smallskip
\caption{\label{falfa} Plot of fractal dimension $f(\alpha)$ against
  the Lipschitz-H\"{o}lder exponent $\alpha$ for SP ($a$) and DSP
  ($b$). Squares represent the square lattice data and triangle
  represent triangular lattice data respectively. For OP, it is a
  single point at $f(\alpha)=\alpha=d_f$ and represented by a
  circle. For DSP and SP, different spectra of $f(\alpha)$s are
  obtained. The spectra on the square and triangle lattice differ
  considerably for DSP whereas for SP they are almost identical.  }
\end{figure}

The associated fractal dimensions $f(\alpha)$ with the measure and the
corresponding Lipschitz-H\"{o}lder exponent $\alpha$ can be obtained
through a Legendre transformation \cite{evertsz-mandel} of the
sequence $\tau(q)$. The Legendre transformation is given below
\begin{equation}
\label{falfae}	
\alpha(q) = -\frac{d\tau(q)}{dq}, \hspace{1cm} f(\alpha)=q\alpha(q)+
\tau(q).
\end{equation}	
The fractal dimensions $f(\alpha)$ are plotted against $\alpha$ in
Fig.\ref{falfa}. The values of $f(\alpha)$ obtained for SP clusters
are shown in Fig.\ref{falfa} ($a$) and that of the DSP clusters are
shown in Fig.\ref{falfa} ($b$). Since in the case of OP, the state
exponent follows a constant gap equation (Eq.\ref{tqdf}) it is
expected that $f(\alpha)$ versus $\alpha$ will be represented by a
point $f(\alpha)= \alpha = d_f$. It is shown by an open circle in
Fig.\ref{falfa}. It has been verified measuring slopes at different
regions of $\tau(q)$ versus $q$ for OP that the slopes remain within
the error bar of the point shown. In the cases of SP and DSP, spectra
of $f(\alpha)$ against $\alpha$ are obtained because $\tau(q)$ has a
non linear dependence on $q$. The symbols square and triangle
correspond to the square lattice and the triangular lattice data
respectively. There are few things to observe. First,
$f_{max}(\alpha)$ corresponds to $d_f$ of the respective
models. Second, the $f(\alpha)$ curves are always $\le d_f$ since the
supports are spanning percolation clusters of fractal dimension
$d_f$. Third, the spectrum of fractal dimensions $f(\alpha)$ for SP
and DSP are found different. It means that not only the mass fractal
dimension $d_f$ is different but also the whole set fractal dimensions
$f(\alpha)$s are different. It is expected because SP and DSP belong
to different universality classes. Fourth, in case of SP, the spectrum
of $f(\alpha)$ for the square lattice is identical with that of the
triangular lattice for lower moments (starting from the same mass
fractal dimension) and slightly different at large positive
moments. In case of DSP, the spectra of $f(\alpha)$ on the two
lattices are considerably different over the full range of positive
moments considered here, starting from two different mass fractal
dimensions. This again confirms the universality of critical exponents
in SP and breakdown of universality of critical exponents in DSP
between square and triangular lattices in two
dimensions\cite{sinha}. Fifth, the values of $f(\alpha_{min})$ for
both SP and DSP clusters are not equal to zero. This means that in
these cases, the rarest regions of measure $\mu_{max}$ distribution
are still fractal in the limit $q\rightarrow \infty$. It is evident
from the spanning clusters configurations given in Fig.\ref{cluster}
that the $\mu_{max}$ distribution appears in small islands allover the
spanning clusters and not as a point distribution. The fractal
dimension of $\mu_{max}$ distribution of SP clusters is found little
higher than that of DSP clusters. This is due to the presence of an
extra directional constraint in the DSP model which takes the growth
of the cluster away from a $\mu_{max}$ point whereas due to pure
rotational constraint in the SP model the probability of growth around
a $\mu_{max}$ point is little higher in comparison to DSP. It can also
be noticed that, in case of DSP, the whole $f(\alpha)$ spectrum is
shifted upward in order to match with the mass fractal dimension of
the spanning DSP clusters on the triangular lattice. However this is
not the case in SP model. Finally, the values of $f(\alpha)$ converges
at a minimum of Lipschitz-H\"{o}lder exponent $\alpha_{min}$. The
Lipschitz-H\"{o}lder exponent $\alpha(\xi)$ is defined as $\mu_\xi=
\delta^{\alpha(\xi)}$ where $\mu_\xi = \mu(\xi+\delta) - \mu(\xi)$ is
the increment in the measure over a length $\xi$ to $\xi+\delta$
\cite{feder}. The $\alpha_{min}$ value corresponds to the minimum of
$\xi$, the length scale associated with $\mu_{max}$ clusters in this
case. It could also be noticed that in both SP and DSP, the values of
$\alpha_{min}$ on the triangular lattice is found little higher than
that of the square lattice. This is due to the fact that the number of
$\mu_{max}$ points are generally higher on the triangular lattice than
that on the square lattice.

\section{Conclusion}

Using the concept of state variable, the percolation transition in
rotationally constrained models is established at the same percolation
threshold $p_c$ defined geometrically. A relative state variable is
defined to study the multifractal aspects of the spanning clusters at
$p=p_c$. It is found that the $q$-moments of the measure is
characterized by a sequence of ``state exponents'' $\tau(q)$ for both
SP and DSP. The existence of a sequence of state exponents confirms
the multifractal character of the distribution of relative state
variable on the infinite clusters of SP and DSP. The OP spanning
clusters are not found multifractal in this measure. Taking Legendre
transformation of $\tau(q)$, different spectra of associated fractal
dimensions $f(\alpha)$ as a function of Lipschitz-H\"{o}lder exponents
$\alpha$ are obtained. The universality of critical exponents in SP
and breakdown of universality in DSP are also confirmed by the
multifractal spectrum of fractal dimensions. The formalism of state
variable is thus found suitable for studying percolation transition
and multifractal aspects of certain percolation models.

\vspace{0.5cm}
\noindent
{\bf Acknowledgment:} SS thanks CSIR, India for financial support.

\newpage

\end{document}